\documentclass[aps,prb,superscriptaddress,twocolumn]{revtex4}
\usepackage{dcolumn}
\usepackage{t1enc}
\usepackage{amsmath}
\usepackage{graphicx}
\usepackage{epsfig}
\usepackage{psfrag}
\usepackage{times}
\usepackage[matha,mathx]{mathabx}

\newcommand {\mb}[1]{\mathbf{#1}}

\begin{document}

\title{Spatial correlation functions and dynamical exponents in very large samples of 4D spin glasses}

\author{Lucas Nicolao}
\thanks{Present address: Departamento de F\'isica, Universidade Federal do Rio Grande do Sul CP 15051, 91501-979, Porto Alegre, Brazil}
\affiliation{Dipartimento di Fisica, Sapienza Universit\`a di Roma, P.le Aldo Moro 2, I-00185 Roma, Italy}

\author{Giorgio Parisi}
\affiliation{Dipartimento di Fisica, INFN -- Sezione di Roma I, IPCF-CNR -- UOS Roma, Sapienza Universit\`a di Roma, P.le Aldo Moro 2, I-00185 Roma, Italy}

\author{Federico Ricci-Tersenghi}
\affiliation{Dipartimento di Fisica, INFN -- Sezione di Roma I, IPCF-CNR -- UOS Roma, Sapienza Universit\`a di Roma, P.le Aldo Moro 2, I-00185 Roma, Italy}

\begin{abstract}
The study of the low temperature phase of spin glass models by means of Monte Carlo simulations is a challenging task, because of the very slow dynamics and the severe finite size effects they show.
By exploiting at the best the capabilities of standard modern CPUs (especially the SSE instructions), we have been able to simulate the four-dimensional (4D) Edwards-Anderson model with Gaussian couplings up to sizes $L=70$ and for times long enough to accurately measure the asymptotic behavior.
By quenching systems of different sizes to the the critical temperature and to temperatures in the whole low temperature phase, we have been able to identify the regime where finite size effects are negligible: $\xi(t) \lesssim L/7$.
Our estimates for the dynamical exponent ($z \simeq 1/T$) and for the replicon exponent ($\alpha \simeq 1.0$ and $T$-independent), that controls the decay of the spatial correlation in the zero-overlap sector, are consistent with the RSB theory, but the latter differs from the theoretically conjectured value.

\end{abstract}

\maketitle

\section{Introduction}

Even though much progress has been made in the past decades, our
comprehension of the underlying nature of the spin glass phase in
finite dimensions faces many open problems \cite{parisi2008}. The two
major scenarios stem from theories that are exact in opposite
dimensional limits. Exact in one dimension, the droplet picture
\cite{bray86, fisher1986} considers only two equilibrium pure states
related by spin-flip symmetry.  In contrast, in the mean-field picture
\cite{MePaVi1987}, replica symmetry is fully broken in a hierarchical
pattern and the many equilibrium pure states are organized in an
ultrametric fashion \cite{MaPaRiRuZu2000, PaRi2000}.

These central features of the mean-field solution survive in finite
dimensions, as the mean-field solution is amenable to computations
down to, and around, the upper critical dimension ($d_u=6$), in the
form of a replicated field theory. It is well known that at and below
the critical temperature there is a massless mode associated with the
breaking of the continuous replica symmetry. Therefore, a spin glass
is always in a critical state due to the coexistence of many
equilibrium states, and the associated overlap-overlap connected
correlation functions decay as a power-law. This Goldstone mode is
called the {\it replicon mode}. In this work we will present results
restricted to the zero overlap sector \footnote{Correlations are
  expected to behave differently in other overlap sectors. Results for
  the replicon exponent for different overlap sectors will appear in a
  forthcoming publication.}. Note that since there is no replica
symmetry breaking in the droplet theory (only two pure states exist),
the overlap-overlap correlation function is not even defined for the zero
overlap sector, and in the $q_{EA}$ sector, where it is defined, it
decays in a standard way.

As it happens in the wider framework of renormalization group for
random systems, spin glasses in zero magnetic field face technical
difficulties, especially for long distances behavior in $6-\epsilon$
dimensions inside the broken phase \cite{Parisi2012}.  At this point,
numerical simulations are very useful and we feel that studying the
four-dimensional case is very important to interpolate between the
field theoretical results above $d_u$ and the tridimensional
case. The latter is the most explored case in large scale numerical
simulations, which is indeed the crucial case, but also very close to
the lower critical dimension (most probably $d_l\simeq 2.5$
\cite{FaPaVi1994,boettcher2005}).

Above $d_u=6$, the strongest infrared behavior among all propagators
is exhibited by the zero overlap replicon, where the associated
overlap-overlap correlations decay as $r^{-\alpha}$ with $\alpha=d-4$
below $T<T_c$ and $\alpha = d-2$ at $T=T_c$\cite{dom98} - verified by
numerical simulations in large-$d$ diluted hypercubes
\cite{FeMaPaSe2010} and in $d=6$~\cite{parisi97}.
This scenario should change for $d<6$, leading to the standard relation
$\alpha=d-2+\eta$ at $T=T_c$ (being $\eta$ the anomalous dimension).
Moreover field theoretical arguments suggest~\cite{domgia06} that
$\alpha=(d-2+\eta)/2$ for $T<T_c$.

The large majority of works devoted to the numerical estimation of the
replicon exponents have been done by Monte Carlo simulations of
tridimensional models. Non-equilibrium methods \cite{Marinari96,
  MaPaRiRu00, Janus08, Janus09} are an alternative to equilibrium
studies \cite{marinari00, marinari01, contucci09}, presenting
compatible values; the latest estimates are $\alpha\simeq 0.4$ at
$T<T_c$ for the zero overlap sector.

Non-equilibrium methods rely on the extrapolation of the dynamical
evolution to allow one to estimate the equilibrium correlation
functions \cite{MaPaRiRu00}, assuming a simple and yet
general Ansatz \cite{Marinari96} for the time dependence in the
overlap-overlap correlation function. However, a very powerful method
based on a set of integral estimators of characteristic length scales
was introduced recently \cite{Janus09}, allowing a more robust and
Ansatz-independent determination of the equilibrium correlation
functions. Although such non-equilibrium methods allow the
study of equilibrium spatial correlations only in a restricted overlap
sector, they benefit from the use of very large lattices and thus having
finite size effects under control, while equilibration of large system
sizes deep in the cold phase is computationally cumbersome.

The only previous numerical determination of the four-dimensional case
used a similar non-equilibrium analysis based on the definition of an
Ansatz, but in rather small system sizes and time windows ($L\le 26$
and $t<60000$ MCS), with a replicon exponent lying in the range
$0.9<\alpha<1.35$ below $T_c$\cite{BeBo02}. In this work we report,
using non-equilibrium methods mentioned above, an almost constant
in temperature replicon exponent $\alpha\simeq 1.0$ and the dynamical
critical exponent (inversely proportional to $T$) with high accuracy,
using unprecedented sizes and time range, where we can observe clearly
the finite-size effects and have them under control.

\section{Model and the correlation function}
\label{model}

We have simulated the Edwards-Anderson model for spin glasses on a
four-dimensional cubic lattice of volume $L^4$ with helicoidal
boundary conditions. The Hamiltonian is
\begin{equation}
\mathcal{H} = -\sum_{\left< \mb{x},\mb{y} \right>} J_{\mb{x} \mb{y}} \sigma_\mb{x} \sigma_\mb{y},
\label{hamilt}
\end{equation}
where $\sigma_{\mb{x}}=\pm 1$ are Ising spin variables located at lattice
position $\mb{x}$ and $J_{\mb{x} \mb{y}}$ are quenched coupling constants joining
pairs of lattice nearest neighbors (denoted by $\left< \mb{x},\mb{y}
\right>$), drawn from a Gaussian probability distribution of zero mean
and unitary variance.

Our study concentrates in the behavior of the correlations
of the replica field $q_{\mb{x}}(t)=\sigma^{(1)}_{\mb{x}}(t)\sigma^{(2)}_{\mb{x}}(t)$:
\begin{equation}
C_4(\mb{r},t) = \overline{ L^{-4} \sum_{\mb{x}} q_{\mb{x}}(t) q_{\mb{x}+\mb{r}}(t)},
\end{equation}
where $\sigma^{(1)}$ and $\sigma^{(2)}$ are two real replicas, meaning
two independent systems evolving with the same couplings. We denote by
$\overline{(\cdots)}$ the average over different realizations of
disorder. In our study we have used the data for $C_4(\mb{r},t)$ measured 
along the directions of the principal axis. We have not
found significant improvement of the statistical errors by averaging
$C_4(\mb{r},t)$ over spherical shells.

We always consider the time evolution of this system quenched from
high temperatures (initial conditions are chosen randomly,
i.e.~$T=\infty$) to a fixed working temperature, below or at the
estimated critical temperature $T_c = 1.805(10)$
\cite{JoKa08}. To mimic the physical evolution we have used the
standard Metropolis dynamics. Since we start with two uncorrelated
replicas, they will typically relax in two orthogonal valleys, so that
the system will remain in the $q=0$ sector for large times
(much longer than times used in this study, due to the very large lattices
we simulate). As a consequence, from dynamics we extract the properties
of the equilibrium $q-q$ correlation in the zero overlap sector.

In order to reach large-scale space and time regimes, we have
developed an optimized code dedicated to the use of SIMD (Single
Instruction Multiple Data) technology, present in practically every
modern CPU, where a single processor is able to perform four
floating-point operations simultaneously. With the help of Streaming
SIMD Extensions (SSE) instructions \cite{intel} we could benefit from
this intrinsic paralelization to perform all the operations involved
in a Monte Carlo simulation (see Ref.~\onlinecite{randomsse} for the linear
congruential pseudo-random number generator and Ref.~\onlinecite{expsse} for the
implementation of the exponential function), updating four
non-interacting spin simultaneously, one in each quarter of the
whole volume.

Since this optimization pushes the processor performances to its
theoretical limit, the overall computation time is strongly affected
by the type and size of the cache memory, as well as its
availability. For a Intel(R) Xeon(R) CPU X5365 at 3.00GHz with a L2
Cache of 4MB, the speedup is of 19 times faster than an equivalently
optimized but non-vectorized code.

We have simulated the off-equilibrium dynamics using linear sizes
ranging from $L=30$ up to $L=70$, for seven temperatures ranging
approximately from $0.3 T_c$ to $0.8 T_c$, plus $T_c$ --- see the
complete set of simulation parameters in Table \ref{datapar}. For each
sample, we saved to disk the couplings and the configurations of
each replica at $2^{i/2}$ MC steps; the analysis was performed offline.

\begin{table}[ht]
\caption{Set of the simulation parameters. Configurations of the 2 replicas at each $2^{i/2}$ MC step where saved to disc, as well as the couplings realization.}
\centering
\vspace{0.2cm}
\begin{tabular}{ l  l c c  }
\hline
$T$ \hspace{1cm} & $L$ \hspace{0.5cm} & \hspace{0.5cm}  $N_{S}$ \hspace{0.5cm}  & \hspace{0.5cm} MC steps \hspace{0.5cm} \\ \hline\hline
1.805& 70 & 40 & $2^{22.5}$ \\
     & 42 & 20 & $2^{23}$ \\
     & 30 & 90 & $2^{23}$ \\  \hline

1.400& 70 & 34 & $2^{19.5}$ \\
     & 42 & 20 & $2^{23}$ \\
     & 30 & 60 & $2^{23}$ \\ \hline

1.263& 70 & 50 & $2^{21}$ \\
     & 54 & 20 & $2^{21.5}$ \\
     & 42 & 36 & $2^{26}$ \\
     & 30 & 211& $2^{23}$ \\ \hline

1.100& 54 & 38 & $2^{21}$ \\
     & 42 & 16 & $2^{22}$ \\
     & 30 & 60 & $2^{23}$ \\ \hline

0.900& 54 & 40 & $2^{20}$ \\
     & 30 & 91 & $2^{23}$ \\ \hline

0.700& 30 & 55 & $2^{23}$ \\ \hline

0.540& 30 & 87 & $2^{23}$ \\ \hline
\end{tabular}
\label{datapar}
\end{table}

\section{Results}
\label{results}

First we analyze the overlap correlation function starting with an
Ansatz known to be a good representation of its functional
form\cite{Marinari96}:
\begin{equation}
C_4(r,t) = \frac{\mathrm{const}}{r^{\alpha}} \exp\left[-\big(r/\xi(t)\big)^{\delta}\right]\;.
\label{ansatz}
\end{equation}
Assuming that the coherence length grows algebraically as
$\xi(t)=Bt^{1/z}$, and following Ref.~\onlinecite{MaPaRiRu00} we perform a fit to
Eq.~(\ref{ansatz}) in two steps. First the time dependence for each fixed
distance is fitted to:
\begin{equation}
-\log C_4(r,t) = A(r) + B(r) t^{-\delta/z} \label{Cfit}\;,
\end{equation}
from which the optimal $\delta/z$ is determined through the minimization of
a spatially averaged $\chi^2$,
\begin{equation}
\hat{\chi}^2=\frac{1}{r_{\mathrm{M}}-2}\sum_{r=3}^{r_{\mathrm{M}}} \chi^2(r)/\mathrm{d.o.f.}(r)\;,
\end{equation}
up to distance $r_{\mathrm{M}}$ where fit to Eq.~(\ref{Cfit}) is still
meaningful. Then we interpolate the coefficients in Eq.(\ref{Cfit}) with
the laws $B(r) = B^{-\delta} r^{\delta}$
and $A(r) = \mathrm{const} + \alpha \log(r)$ at the optimal $\delta/z$
to obtain the best estimates for the exponents $\alpha$, $\delta$ and $z$.
In general the $\hat{\chi}^2$ minimization has been performed for
distances $r\ge 3$. However in some cases we found necessary to use
shorter distances for the estimate of $\alpha$ and eventually a
quadratic term $C(r) t^{-2\delta/z}$ in Eq.~(\ref{Cfit}) has been added.

\begin{table}[ht]
\caption{Best fitting parameters and corresponding fitting ranges obtained by
interpolating spatial correlation functions with the Ansatz in Eq.~(\ref{ansatz}).
In the cases marked by $^*$ a second order term, $t^{-2\delta/z}$, has been added to
Eq.~(\ref{Cfit}).}
\centering
\begin{tabular}{ccccccccc}
$T$ & $L$ & $[r_{i},r_{f}]$ & $\delta$ & $B^{-\delta}$ & $z$ & $[r_{i},r_{f}]$ & $\alpha$ \\ \hline
1.400& 70& [3,11]& 1.47(4)&  1.00(5)&	 6.18(32)&  [1,7]&  1.10(6)$\,\,\,$\\
     & 42&  [3,8]& 1.49(5)&  0.98(9)&	 6.26(48)&  [1,8]&  1.11(4)$^*$\\
1.263& 70& [3,12]& 1.48(3)&  0.96(2)&	 6.79(17)&  [1,5]&  1.07(2)$^*$\\
     & 42& [3,12]& 1.52(8)&  0.78(2)&	 7.68(37)&  [1,6]&  1.05(3)$^*$\\
     & 30& [3,10]& 1.48(4)&  0.92(4)&	 6.91(37)&  [1,6]&  1.09(2)$^*$\\
1.100& 54&  [3,9]& 1.51(4)&  0.88(6)& 7.82(74)&  [1,4]&  1.04(6)$\,\,\,$\\
     & 30&  [3,9]& 1.52(8)&  0.76(6)& 8.9(1.5)&  [1,7]&  1.03(7)$^*$\\
0.900& 30&  [3,8]& 1.51(7)&  0.77(6)& 10.1(2.2)& [1,4]&  1.05(7)$^*$\\
0.700& 30&  [2,6]& 1.50(4)&  0.86(4)& 11.0(1.6)& [1,3]&  1.08(11)\\
0.540& 30&  [2,6]& 1.54(6)&  0.81(3)& 13.0(2.8)& [1,3]&  1.09(15)\\
\end{tabular}
\label{Tabfits}
\end{table}

Table \ref{Tabfits} summarizes the best estimates obtained with this procedure.
With high accuracy the exponent of the stretched exponential is constant
throughout the low temperature phase with value $\delta = 1.50(1)$.
The dynamical critical exponent dependence on the temperature is very well
described by the law $z \simeq 8.9(2)/T$ and
the replicon exponent is nearly constant with average $\alpha \simeq 1.06(6)$.

This fitting method provides a reliable estimation for the exponents and
it is certainly a robust way to get the $t \rightarrow \infty$ limit
through a global fit \cite{BeBo02} to the Ansatz in Eq.~(\ref{ansatz}).
Still it suffers from some drawbacks: it is
Ansatz dependent and some technical aspects are not perfectly under control.
For example, in order to ensure a fair spatial average of the $\hat{\chi}^2$
in the first step of the procedure, short distances have to be carefully
selected in order not to dominate over the longer distances. It
can be difficult to precise whether this time window sits between an
initial fast transient dynamics and a near equilibrium dynamics when
$\xi(t) \gg r$.

To overcome these issues, we move to an Ansatz-independent method to
estimate $z$ and $\alpha$, through a set of integrals of the
correlation function in the form
\begin{equation}
I_k(t) \equiv \int_0^{L/2} dr\, r^k\, C_4(r,t)\;,
\label{intk}
\end{equation}
and, since we expect a scaling form $C_4(\mb{r},t)\sim r^{-\alpha} f(\frac{r}{\xi})$, then
\begin{equation}
I_k(t) \, \propto \,\, \xi(t)^{k+1-\alpha}\;,
\label{Ik-xi}
\end{equation}
and we can estimate the coherence length as
\begin{equation}
\xi_{k,k+1} (t) \equiv \frac{I_{k+1} (t)}{I_{k} (t)}\;.
\label{xikkp1}
\end{equation}

From Eqs.~(\ref{Ik-xi},\ref{xikkp1}) it is possible to estimate the replicon
exponent $\alpha$.
As matter of example, in case the correlation function is given exactly by Eq.(\ref{ansatz}) with $\alpha=1$ and $\delta=1.5$ we have that
\begin{equation}
\xi_{2,3}(t) = \xi(t) / \Gamma(4/3) = 1.12\;\xi(t)\;,
\label{twoXi}
\end{equation}
with the constant weakly depending on $\alpha$ and $\delta$.

In order to evaluate the integrals above, we adopt the same procedure
used in Refs.~\onlinecite{Janus09,Yllanes11}, introducing a self-consistent
integration cutoff at a distance where the correlation function first becomes
less than $X$ times its statistical error, with $X=5$ but for
$T=T_c$ ($X=7$) and $T=0.9$ ($X=4$). As this method alleviates
the integrals from the wide fluctuations of the non self-averaging
tails, it also induces a systematic error. To avoid such a systematic error,
we estimate the contribution of the tail by performing a fit to the Ansatz
in Eq.~(\ref{ansatz}) using our previous estimates of $\delta\simeq 1.5$ and
$\alpha\simeq 1.0$ for $T<T_c$ and the best previous estimate\cite{JoKa08,PaRiRu96}
$\alpha \simeq 1.7$ for $T=T_c$. The fit is performed in the range
$[3,\min(r_{\mathrm{max}},r_{\mathrm{cutoff}})]$ with $r_{\mathrm{max}}=10$,
$12$ and $15$ for $L=30$, $42$ and $L\ge 54$ respectively, and is used to
estimate the integral in the range $\left[r_{\mathrm{cutoff}},L/2\right]$.

\begin{figure}[t]
\includegraphics[width=\columnwidth]{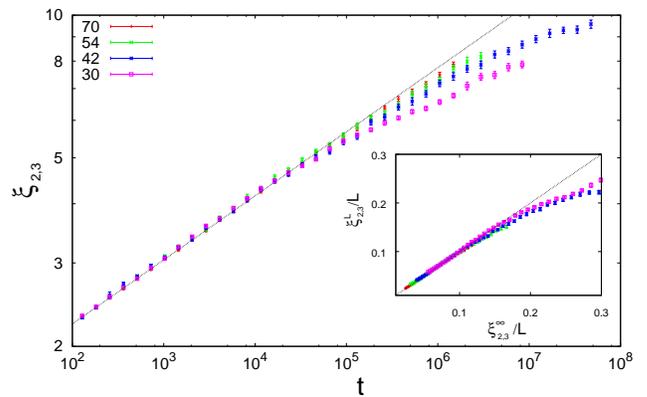}\\
\caption{Time dependence of the coherence length for $T=1.263$ and all
  system sizes simulated. The line accounts for the power-law fit to
  $L=70$ data in the range $\xi_{2,3}\in [3,5]$, that
  corresponds to $\xi^{\infty}_{2,3} (t) \equiv 1.20(1) \,
  t^{1/7.44(10)}$. The inset shows the scaling function of the
  coherence length, evidencing the point where finite-size effects
  become important, $L \lesssim 6.5\,\xi_{2,3}(t)$.}
\label{x1.26-Ls}
\end{figure}

In Fig.~\ref{x1.26-Ls} we report the time dependence of
$\xi_{2,3}(t)$ for various system sizes at $T=1.263$.
In the early stage of the dynamics, in general for $\xi(t) \lesssim 3$,
spatial correlation functions show the effects due to the lattice
discretization and the growth of $\xi(t)$ shows some pre-asymptotic
behavior.
In the inset of Fig.~\ref{x1.26-Ls} we show that finite-size effects
come to play when $L \lesssim 6.5\, \xi_{2,3}(t) \simeq 7 \xi(t)$,
much earlier  than the standard expectation, $\xi(t) \simeq L/2$.
On top of the finite size effects, we also observe some deviations
due to the uncertainty in estimating the contribution of the tail.
As a consequence the estimation of the $z$ exponent from the fit
$\xi_{2,3}(t)\,\propto\, t^{1/z}$ must be restricted to a time window
that excludes both the short time dynamics (affected by lattice
discretization) and the very long time dynamics, even for the largest volumes.

\begin{figure}[t]
\includegraphics[width=\columnwidth]{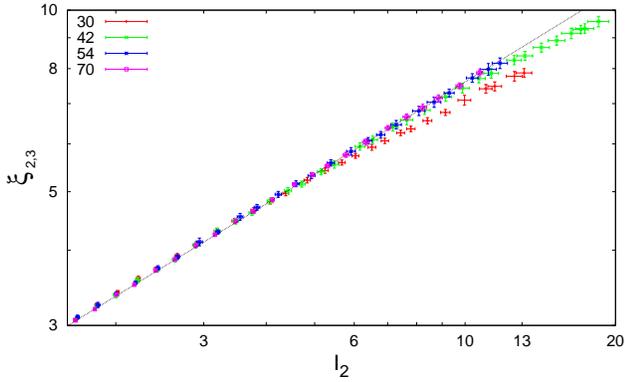}
\caption{Data for $T=1.263$; different sizes have different number
  of data point, according to the length of the simulation shown in
  Table~\ref{datapar}.
  The line corresponds to $\alpha=1.033$ obtained from fitting
  the $L=70$ data. Deviations from this line are finite size effects.
  Although the simulations for $L=42$ have been
  run longer for a factor 32, it is clear that the last part of the
  simulation is affected by strong finite size effects and practically
  useless for the estimation of the exponent.}
\label{xi_IT1.263}
\end{figure}

Fortunately enough, all integrals $I_k$ experience the same inaccuracy
in the extrapolation of the tail contribution, so that these errors
compensate each other in the relation between $\xi_{k,k+1}$ and $I_k$.
In Fig.~\ref{xi_IT1.263} we clearly identify the finite size effects, but
there are no other systematic errors due to the tail extrapolation.
Since $I_2\,\propto\,\xi_{k,k+1}^{3-\alpha}$ we can extract the replicon
exponent from a direct fit to the relation (\ref{Ik-xi}), without
discarding late time data for the largest sizes.
We use a standard method for linear fits with errors in both
coordinates \cite{nr}.

In principle, we could have used other values for $k$ and $m$ in
$\xi_{k,k+1}$ and $I_m$ appearing in the relation (\ref{Ik-xi}). The
choice for $\xi_{2,3}$ and $I_2$ is justified because it brings the
highest amount of points for the fits of $\xi_{k,k+1}(I_k)$ to a
power-law, namely from 20\% to 50\% less discarded short-time data.
To keep consistency with the $\alpha$ estimation, we have chosen
$\xi_{2,3}$ for the estimate of exponent $z$.

\begin{figure}[t]
\includegraphics[width=\columnwidth]{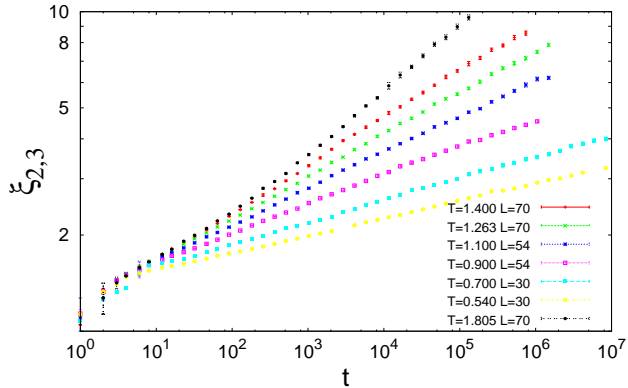}
\caption{Time dependence of the coherence length $\xi_{2,3}$ for the largest size $L=70$ and several temperatures, ranging from the critical one $T_c=1.805$ down to very low temperatures.}
\label{xi_t}
\end{figure}

\begin{figure}[t]
\includegraphics[width=\columnwidth]{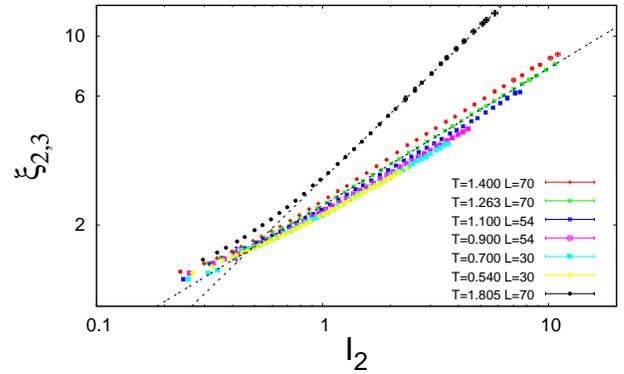}
\caption{Data for $L=70$ and several temperatures.
  The relation $\xi_{2,3}\,\propto\,I_2^{1/(3-\alpha)}$ provides, at large
  times, a reliable estimate for the replicon exponent $\alpha$.
  Lines correspond to $\alpha=1.02$ for $T=1.263$ and $\alpha=1.77$ for
  $T=T_c=1.805$. The latter exponent
  corresponds to $\eta=-0.23(1)$, which is compatible with the latest
  estimate using equilibrium finite size scaling analysis\cite{JoKa08}:
  $\eta=-0.275(25)$.}
\label{xi_I}
\end{figure}

\begin{table}[t]
\caption{Best estimates for the dynamical and replicon exponents from
  the fitting of the data shown in Fig.~\ref{xi_t} and Fig.~\ref{xi_I}.
  The fits are performed on the range $[\xi_{\mathrm{min}},\xi_{\mathrm{max}}]$
  for $z$, and with $\xi_{2,3} \ge \xi_{\mathrm{min}}$ in the case of $\alpha$.
  Errors are obtained through jackknife methods.
}
\centering
\begin{tabular}{cccccc}
$T$ & $[\xi_{\mathrm{min}},\xi_{\mathrm{max}}]$ & $z$ & $\chi^2_{\xi}$/d.o.f. & $\alpha$ & $\chi^2_{I_2(\xi)}$/d.o.f. \\ \hline 
1.805&  [3.5,13.0]&  4.95(04)& 19/21 &  1.766(03)& 3.4/21\\
1.400&   [3.5,8.0]&  6.86(14)& 5.1/16&  1.055(19)&  1.5/18\\
&        [4.0,8.0]&  6.89(17)& 3.3/14&  1.050(22)&  0.7/16\\
1.263&   [3.0,5.0]&  7.44(10)& 6.2/11&  1.020(10)&  3.4/20\\
&        [3.0,5.5]&  7.45(08)& 7.1/13&           &        \\
&        [3.5,5.5]&  7.46(10)& 3.4/10&  1.015(13)&  2.1/17\\
1.100&   [3.0,5.5]&  9.15(15)& 4.0/15&  0.996(19)&  7.8/20\\
&        [3.5,5.5]&  9.21(20)& 3.4/11&  1.024(29)&  5.5/16\\
&        [3.0,6.5]&  9.19(16)& 7.8/20&           &        \\
&        [3.5,6.5]&  9.23(27)& 6.6/16&           &        \\
0.900&   [2.7,4.0]& 11.32(33)& 6.6/14&  0.909(31)& 1.8/19 \\
&        [3.0,4.0]& 11.50(46)& 5.9/10&  0.921(42)& 1.4/15 \\
0.700&   [2.7,4.5]& 15.31(67)& 2.3/18&  0.900(36)&  1.1/18\\
&        [3.0,4.5]& 15.47(77)& 1.3/14&  0.923(40)&  0.4/14\\
0.540&   [2.3,3.3]& 17.9(1.3)& 12/17 &  0.896(39)&  2.0/17\\
&        [2.6,3.3]& 19.6(2.1)& 5.4/11&  0.86(20) &  1.8/11\\
\end{tabular}
\label{data}
\end{table}

\begin{figure}[t]
\includegraphics[width=\columnwidth]{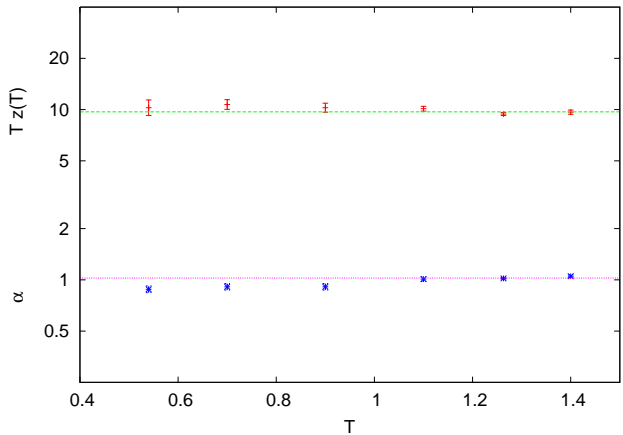}
\caption{Best estimates for $z(T)$, multiplied by $T$, (above) and $\alpha$ (below)
taken from Table~\ref{data}. Lines are best fits: $T z(T) = 9.7(2)$ and $\alpha=1.025(9)$.}
\label{newPlot}
\end{figure}

Our main results are summarized in Fig.~\ref{xi_t} and Fig.~\ref{xi_I}
where $\xi_{2,3}(t)$ is shown as a function of time and $I_2(t)$, respectively,
for several temperatures.
From data in Fig.~\ref{xi_t} we estimate the dynamical exponent $z$ by fitting to a
power law in the range $\xi\in[\xi_{\mathrm{min}},\xi_{\mathrm{max}}]$,
and from the data in Fig.~\ref{xi_I} we get the replicon exponent $\alpha$ fitting
in the range $\xi \ge \xi_{\mathrm{min}}$.
It is immediately clear from the observation that all data with $T< T_c$
in Fig.~\ref{xi_I} become parallel at large times that $\alpha$ is roughly
constant in the low temperature phase.
Our best estimates for the exponents $z$ and $\alpha$ are reported in Table \ref{data},
together with the fitting ranges, the corresponding $\chi^2$ and
number of degree of freedom (d.o.f.).
In Fig.~\ref{newPlot} we plot the best estimate for the exponents $z$ and $\alpha$
in a way that makes evident that $z(T) \simeq 9.7(2)/T$ and $\alpha\simeq 1.025(9)$
for $T<T_c$.
Actually the best value for $\alpha$ has been estimated only from data in the range
$T\in[1,1.5]$, because for lower temperatures we observe a systematic decrease in the
$\alpha$ value that we explain as follows.
From data plotted in Fig.~\ref{xi_I} we see that systems at the lowest temperatures
are still approaching the asymptotic dynamics; so, it is likely that the small drift
of the exponents for these low temperatures does not reflect a real change, but
rather a preasymtotic effect due to a not large enough value of $\xi(t)$.
Moreover in the $T=0$ limit we could have in principle a different exponent and we
are maybe observing the beginning of the crossover region.
Please note that the relatively small sizes used at lowest temperatures
(see Table~\ref{datapar}) do not induce any finite size effect, because the
growth of $\xi(t)$ is extremely slow and barely reaches $\xi \sim 4 < L/7$.

\section{Conclusions}

We have performed an extensive numerical study of the 4D Gaussian EA model with the aim of measuring the dynamical and the replicon exponents.
We have used very large system sizes (up to $70^4$), which were never used before.
These huge sizes are required to overcome finite size effects, which appear when the coherence length $\xi(t)$ is of the order of $1/7$ of the system size.
The values of the replicon exponent $\alpha$, controlling the spatial decay of
the correlation function, $C_4(r,t) \sim r^{-\alpha}$, are roughly constant in
the low temperature region (apart from some pre-asymptotic effects at very low
temperatures). Our final conservative estimate is $\alpha=1.03(2)$, which is not
far, but definitely different from the one conjectured by the field theoretical
arguments based on the analysis of the first order in the $\epsilon=6-d$ expansion,
$(d-2+\eta)/2 = 0.883$.
We also confirm that the dynamical exponent $z(T)$ is inversely proportional
to the temperature in the entire temperature range we studied.

\appendix
\section{A closer look on integral estimators}
\label{appA}

\begin{figure}[t]
\includegraphics[width=\columnwidth]{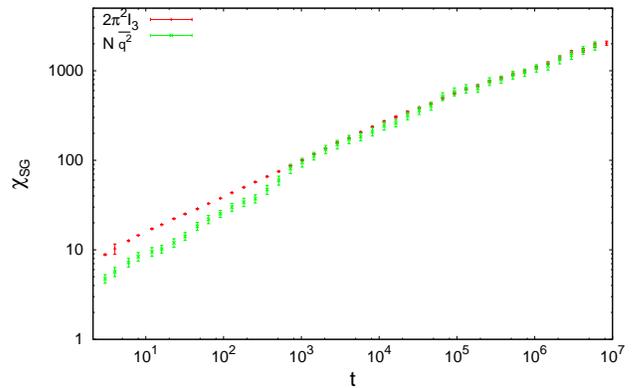}
\caption{We check the reliability of the integral estimator $I_3$
  comparing it to the SG susceptibility for the $L=30$ and $T=1.263$
  case, where we have the largest number of samples (211).}
\label{CSG}
\end{figure}

In this appendix we explore in some detail the properties of the
integrals $I_k$ defined by Eq.~(\ref{intk}) as well as the related estimators
for the coherence length defined by Eq.~(\ref{xikkp1}). Note that the
spin glass susceptibility is given by $\chi^{\mathrm{SG}}(t) = 2\pi^2 I_3(t)$.
This relation offers a check for the correctness of the computation of the integrals.
Since $\chi^{\mathrm{SG}} = N \overline{q^2}$ is a non self-averaging quantity
we do such a comparison for the case in which we dispose the largest set of
samples, see Fig.~\ref{CSG}.

\begin{figure}[t]
\includegraphics[width=\columnwidth]{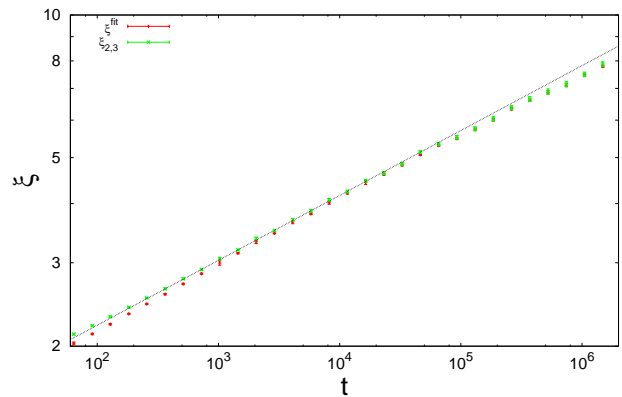}
\caption{Comparison between the integral estimator for the coherence length,
  $\xi_{2,3}$, and the $\xi^{\mathrm{fit}}$ obtained by fitting to the
  Ansatz in Eq.~(\ref{ansatz}). Data are for $L=70$, $T=1.263$ and the
  values of $\xi^{\mathrm{fit}}$ have been multiplied by a factor
  $1/\Gamma(4/3)=1.12$ in order to be equal to $\xi_{2,3}$
  in case Ansatz in Eq.~(\ref{ansatz}) is asymptotically exact.
  Fitting these data in the range $3<\xi<5$ we get $z(\xi_{2,3})=7.4(1)$
  and $z(\xi^{\mathrm{fit}}) = 7.23(7)$. We draw a line of slope 1/7.3 as a
  guide for the eyes.}
\label{xis}
\end{figure}

A comparison between the integral estimator $\xi_{2,3}$ for the coherence
length and another estimator $\xi^{\mathrm{fit}}$ can be seen in Fig.~\ref{xis}.
The latter is obtained by fitting correlation functions $C_4(r,t)$ with the
Ansatz in Eq.~(\ref{ansatz}) with $\alpha=1.0$ and $\delta=1.5$.
These two estimators are in agreement with each other, once normalized according
to Eq.~(\ref{twoXi}). A closer inspection reveals the deviation
of $\xi_{2,3}$ at larger times due to the badness of the estimation of
the tail contribution, so that a secure range for a fit to obtain $z$
is $\xi_{2,3}\in [3,5]$, though compatible results are obtained in a
wider time window, as can be seen in Table~\ref{data}.

\begin{acknowledgments}
L.~Nicolao would like to thank David Yllanes for the enlightening
discussions. The authors acknowledge financial support from the
European Research Council (grant agreement no.~247328) and from the
Italian Research Ministry (FIRB project no.~RBFR086NN1). L.~Nicolao
also acknowledges the initial financial support from CNPq (Conselho
Nacional de Desenvolvimento Cient\'ifico e Tecnol\'ogico), Brazil.

\end{acknowledgments}

\bibliography{4dEAG}

\end{document}